\documentclass[twocolumn]{aastex62}

\usepackage{color}
\usepackage{amsmath}

\newcommand{\Teff}{\mbox{$T_\mathrm{eff}$}}
\newcommand{\Mjup}{\mbox{$M_\mathrm{Jup}$}}

\newcommand{\fsed}{\mbox{$f_{\rm sed}$}}
\newcommand{\fhole}{\mbox{$f_{\rm h}$}}

\shorttitle{Strong Variability of the Young Cloudy L Dwarf Companion VHS J1256--1257 b}
\shortauthors{Bowler, Zhou, et al.}

\begin{document}

\title{Strong Near-Infrared Spectral Variability of the Young Cloudy L Dwarf Companion VHS J1256--1257 b}

\correspondingauthor{Brendan P. Bowler}
\email{bpbowler@astro.as.utexas.edu}

\author[0000-0003-2649-2288]{Brendan P. Bowler}
\affiliation{Department of Astronomy, The University of Texas at Austin, Austin, TX 78712, USA}

\author{Yifan Zhou}
\affiliation{McDonald Observatory and the Department of Astronomy, The University of Texas at Austin, Austin, TX 78712, USA}
\affiliation{McDonald Prize Fellow}

\author{Caroline V. Morley}
\affiliation{Department of Astronomy, The University of Texas at Austin, Austin, TX 78712, USA}

\author{Tiffany Kataria}
\affiliation{Jet Propulsion Laboratory, California Institute of Technology, 4800 Oak Grove Drive, Pasadena, CA, USA}

\author{Marta L. Bryan}
\affiliation{Department of Astronomy, 501 Campbell Hall, University of California Berkeley, Berkeley, CA 94720-3411, USA}

\author{Bj\"{o}rn Benneke}
\affiliation{University of Montreal, Montreal, QC, H3T 1J4, Canada}

\author{Konstantin Batygin}
\affiliation{Division of Geological and Planetary Sciences, California Institute of Technology, Pasadena, CA 91125 USA}

\begin{abstract}

Rotationally-modulated variability of brown dwarfs and giant planets provides unique information about their
surface brightness inhomogeneities, atmospheric circulation, cloud evolution, 
vertical atmospheric structure, and rotational angular momentum.  
We report results from \emph{Hubble Space Telescope}/Wide Field Camera 3 
near-infrared time-series spectroscopic observations of three companions with masses 
in or near the planetary regime:
VHS J125601.92--125723.9 b, GSC 6214--210 B, and ROXs 42 B b. 
VHS J1256--1257 b exhibits strong total intensity and spectral variability 
with a brightness difference of 19.3\% between 1.1--1.7 $\mu$m over 8.5 hours  
and even higher variability at the 24.7\% level at 1.27~$\mu$m.  
The light curve of VHS J1256--1257 b continues to rise at the end of the observing sequence so these 
values represent lower limits on the full variability amplitude at this epoch. 
This observed variability rivals (and may surpass) the most variable brown dwarf currently known,
2MASS J21392676+0220226.
The implied rotation period of VHS J1256--1257 b is $\approx$21--24 hr assuming sinusoidal modulations, 
which is unusually long for substellar objects.
No significant variability is evident 
in the light curves of GSC 6214--210 B ($<$1.2\%) and ROXs 42 B b ($<$15.6\%).
With a spectral type of L7, an especially red spectrum, and a young age, 
VHS J1256--1257 b reinforces emerging patterns between high variability amplitude, 
low surface gravity, and evolutionary phase near the L/T transition. \\
\end{abstract}

\section{Introduction}{\label{sec:intro}}

Time series photometry and spectroscopy of brown dwarfs 
has opened up a new
window into the physical properties and atmospheric structure of substellar objects.
Over the past decade, high precision infrared monitoring programs 
have demonstrated that most brown dwarfs 
are variable at the 0.2--5\% level between 1 to 5 $\mu$m,
especially those spanning the L and T spectral classes between $\approx$500--2500~K
(e.g., \citealt{Artigau:2009bk}; \citealt{Apai:2013fn}; \citealt{Radigan:2014dj}; \citealt{Metchev:2015dr}).
There is now abundant evidence that the primary source of this variability  
is from rotationally-modulated surface features in the form of
evolving heterogeneous  condensate clouds (e.g., 
\citealt{Showman:2013is}; \citealt{Crossfield:2014cy}), 
at least among low-temperature late-L and T-type brown dwarfs,
analogous to Jupiter's latitudinally  
banded structure where gaps in the cloud deck result in bright regions at infrared wavelengths 
(e.g., \citealt{Antunano:2019aa}; \citealt{Ge:2019bh}).
These patchy clouds produce hot and cold spots which, coupled with rotation, 
result in periodic disk-integrated brightness variations that evolve in amplitude, phase,
and wavelength (\citealt{Buenzli:2012gd}; \citealt{Apai:2017dua}).

Most of these monitoring campaigns have focused on old (several Gyr) brown dwarfs in the field.
Recently there has been increased interest in exploring the variability properties of both 
young brown dwarfs 
and giant planets found with high-contrast imaging 
to examine the influence of surface gravity on variability properties.
For example, \citet{Biller:2015bu} found that the $\approx$8~\Mjup \ object PSO J318.5--22 
exhibits strong variability at the 10\% level in $J$ band.
\citet{BenWPLew:2016kg} report that the young isolated brown dwarf WISE~J0047+6803 
is highly variable from 1.1--1.7~$\mu$m, with peak-to-peak brightness changes as high as 8\%.
Variability has also been confidently observed in a growing number of companions
at or below the deuterium-burning limit ($\approx$13~\Mjup): 
2M1207b (\citealt{Zhou:2016gc}), 
Ross 458 C (\citealt{Manjavacas:2019aa}), 
HD 203030 B (\citealt{Miles-Paez:2019aa}), 
HN Peg B (\citealt{Zhou:2018ij}), 
2M0122--2439 b (\citealt{Zhou:2019fk}), and
GU Psc b (\citealt{Naud:2017do}; \citealt{Lew:2019aa}). 

Several statistical patterns are emerging from variability surveys of substellar objects.
There is evidence that low-gravity brown dwarfs have higher variability amplitudes
compared to their higher-gravity counterparts (\citealt{Metchev:2015dr}).
\citet{Vos:2018aa} 
found that young L dwarfs have higher intrinsic rates of
variability compared to field brown dwarfs with 98\% confidence, an indication that low surface gravity
plays an important role in shaping the spatial distribution and physical properties of condensate clouds.
Viewing geometry also appears to impact these observational signatures;
brightness changes are most strongly attenuated for brown dwarfs 
observed at low inclinations (closer to pole-on orientations; \citealt{Vos:2017ft}).


\begin{figure*}
  \vskip -1.2 in
  \hskip -.7 in
  \resizebox{8in}{!}{\includegraphics{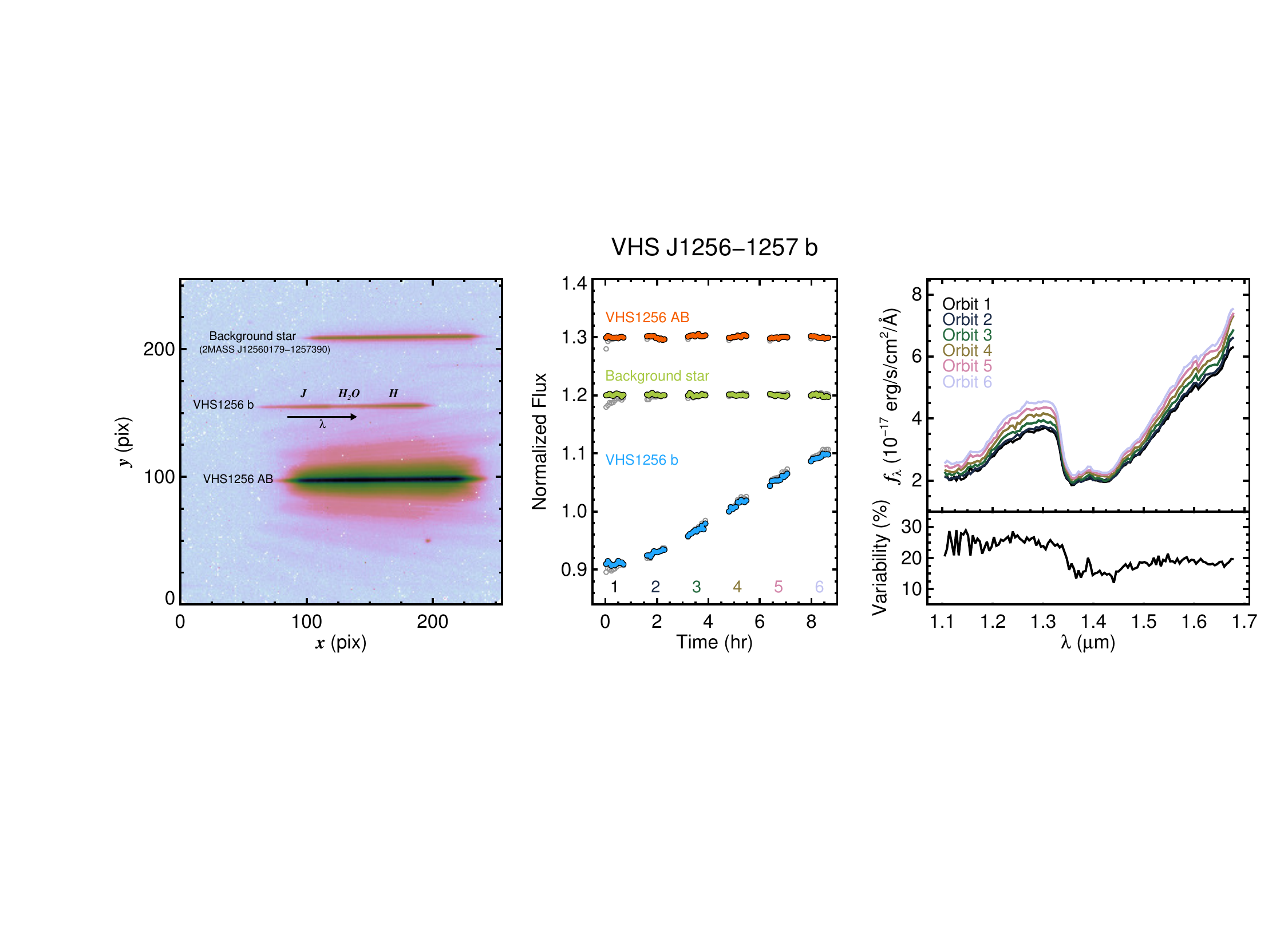}}
  \vskip -1.7 in
  \caption{\emph{HST} time-series grism observations of VHS J1256--1257 b.  \emph{Left}: Grism
  image of VHS J1256--1257 AB, VHS J1256--1257 b, and the nearby background star 
  2MASS J12560179--1257390.  Wavelength increases to the right along the dispersion direction. 
  \emph{Middle}: Extracted and normalized light curves of all three objects from 1.1--1.68~$\mu$m.  
  VHS J1256--1257 AB and the nearby background star are approximately constant over all 
  6 orbits ($\approx$8.5 hours), whereas VHS J1256--1257 b clearly 
  shows a substantial increase in brightness.
  Gray open circles represent the uncorrected photometry;
  colored symbols show the normalized flux after correcting for ramp effects 
  using our RECTE pipeline (see Section~\ref{sec:reduc} for details).
  Orbits are labeled at the bottom of the panel.  
  \emph{Right}: Median spectra for each orbit (top panel) and relative spectral variability (max/min -- 1; bottom panel).
  Variability is strongest in $J$ band and weakest in the 1.4~$\mu$m water band.  
  \label{fig:grism1} } 
\end{figure*}

Extending variability studies to directly imaged exoplanets is challenging because of their 
high contrasts and close separations.  Over the past decade, a steadily growing population of giant planets 
located at unexpectedly wide separations of tens to hundreds of AU from their host stars has been identified.
Their relatively wide separations ($>$100~AU) and modest contrasts ($<$10 mag) 
make them amenable to detailed photometric and spectroscopic observations.
Here we present a \emph{Hubble Space Telescope} ($HST$) program to obtain
time-series spectroscopic light curves of the three 
low-mass companions 
VHS J125601.92--125723.9 b (herinafter VHS~J1256--1257 b), GSC 6214--210~B, and ROXs 42~B~b.
Our Wide-Field Camera 3 (WFC3) \emph{G141} grism observations 
span 1.1--1.7~$\mu$m, enabling simultaneous 
monitoring at $J$ band, $H$ band, and over the 1.4~$\mu$m water feature.
None of these companions have previously been monitored for variability.

VHS~J1256--1257 b is an unusually red L7 companion orbiting a young binary brown dwarf at 8$\farcs$1  
(\citealt{Gauza:2015fw}; \citealt{Stone:2016fz}).
Depending on the system's age and distance, which remain poorly constrained,
the companion's luminosity implies a lower mass limit between
11--26~\Mjup \ (\citealt{Rich:2016dm}).
The $L$-band spectrum of VHS~J1256--1257 b from \citet{Miles:2018dj}
shows signs of weak methane absorption and thick photospheric clouds.
GSC 6214--210 B is an M9.5 companion with a mass of $\approx$14~\Mjup \  orbiting at 2$\farcs$2 (240 AU)
from its young ($\approx$17~Myr) Sun-like host star 
(\citealt{Ireland:2011id}; \citealt{Bowler:2014dk}; \citealt{Pearce:2019iv}).
ROXs 42 B b has a mass of $\approx$10~\Mjup \ and a near-infrared spectral type of L1
(\citealt{Kraus:2014tl}; \citealt{Currie:2014gp}; \citealt{Bowler:2014dk}).
It is the most challenging
target in our sample to observe with $HST$ because of its close angular separation 
to its host star (1$\farcs$2, or 140 AU).
All three companions have rotational broadening measurements from 
high-resolution near-infrared spectroscopy (\citealt{Bryan:2018hd}).

This paper is organized as follows.  In Section~\ref{sec:obs} we describe the $HST$ observations,
spectral extraction, and corrections for detector ramp effects.
In Section~\ref{sec:results} we discuss the companion light curves and interpretation of
variability observed in VHS~J1256--1257 b.  Our observations and
results are summarized in Section~\ref{sec:summary}.

\section{Observations}{\label{sec:obs}}

Time-series spectroscopic monitoring was carried out with \emph{HST}'s WFC3/IR camera
with the $G141$ near-infrared grism
($\lambda$$\lambda$=1.1--1.7~$\mu$m; $\lambda$/$\delta \lambda$$\approx$130) on UT 2018 March 5--6,  UT 2018 March 14--15, and UT 2018 June 14
for VHS~J1256--1257 b, ROXs 42~B~b, and GSC 6214--210~B, respectively (\emph{HST} Proposal GO-15197).
The spectral grasp of our observations samples a range of atmospheric pressure levels
and prominent molecular absorption features, most notably 
the water feature centered at 1.4~$\mu$m and methane beyond 1.6~$\mu$m.
Our program was designed to span
8.5 hr over six contiguous orbits for each target
with unobservable gaps of about 45 min during each orbit.
Observations were carried out in nominal stare mode 
with the telescope roll angle constraints oriented so that the grism dispersion
direction was approximately orthogonal to the star-planet position angle.

At the start of each orbit, several direct images were taken in the $F132N$ filter to 
determine the location of the companion and to calibrate the wavelength solution.
The 256$\times$256 subarray ($\approx$30$''$$\times$30$''$ field of view) was read out to 
reduce buffer dumps and improve readout time efficiency.
Exposure times and number of non-destructive reads were chosen to avoid
saturating the host star at the location of the companion and to ensure 
adequate signal-to-noise ratio of the companion spectrum.
No dithering was carried out to mitigate flat fielding errors.

We acquired 11 grism images of VHS~J1256--1257 b per orbit for a total of 66 frames,
each with an exposure time of 224 s and a SPARS25, NSAMP = 11 readout sequence.
For ROXs 42 B b, 21 grism images were taken for the first orbit followed by 22 for orbits 2--6,
totaling 131 images altogether.  The integration time was 103 s per image and the SPARS10, NSAMP = 15
sequence was chosen.  137 grism images were obtained for GSC 6214--210 B: 
22 were acquired in orbit 1 followed by 23 frames during orbits 2--6.
Each had an integration time of 103 s with SPARS10, NSAMP = 15.

\subsection{Data Reduction}\label{sec:reduc}

We adopt different strategies to extract the grism spectra of our three targets. 
For VHS J1256--1257 b, the contamination from the host star spectral trace at the 
location of the companion (a separation of 57.7 pixels) is negligible:
the average flux level at the location of VHS J1256--1257 b is about 0.23 e$^{-}$ s$^{-1}$ pixel$^{-1}$, 
which is less than 20\%  of the average sky background. We therefore do not carry out point spread function 
(PSF) subtraction for this system.
For the other two targets, the contamination from their host stars is much 
more significant and requires PSF subtraction in the dispersion direction
to recover the companions.
Below we describe details of the data reduction and spectral extraction procedure for each system in this program.

\subsubsection{VHS~J1256--1257 b}
Basic image reduction is carried out with STScI's \texttt{calwfc3} pipeline, which 
performs bias correction, linearity correction, dark subtraction, unit/gain conversion, and cosmic ray identification 
with up-the-ramp fitting.  We extract individual spectra from the 
\emph{flt} images with our custom WFC3/IR time-resolved spectroscopic pipeline, which 
makes use of the \texttt{aXe} slitless spectral extraction package (\citealt{Kummel:2009vi}) and has been 
employed in multiple studies to measure rotational modulations of brown dwarfs 
(e.g., \citealt{Buenzli:2012gd}; \citealt{Apai:2013fn}; \citealt{Buenzli:2014bya}; \citealt{Zhou:2018ij}).
As part of this pipeline, we first expand the $256\times256$ subarray to the $1014\times1014$ full-array images so that they are compatible with \texttt{aXe}. We then identify bad pixels with data quality flag 4 (indicating a bad detector pixel), 16 (hot pixel), 32 (unstable response), and 256 (full-well saturation) and correct them with bi-linear interpolation. We also search and correct for pixels that are affected by cosmic rays but not corrected by the up-the-ramp fit. These pixels are identified by comparing the time sequence of each individual pixel to its median-filtered light curve (in a window size of 5 pixels) and select $7\sigma$ outliers. These pixels are then corrected through bi-linear interpolation with adjacent unaffected pixels. We develop and implement our own sky subtraction routine following \citet{Brammer:2015aa} to incorporate and optimize recently-updated multi-component WFC3 master sky images, and thus run the \texttt{axeprep} task with the ``background'' option turned off. 
Then the \texttt{axecore} task is executed with a 4-pixel radius window to extract spectra.


\begin{figure*}
  \vskip -1.6 in
  \hskip -.4 in
  \resizebox{8.in}{!}{\includegraphics{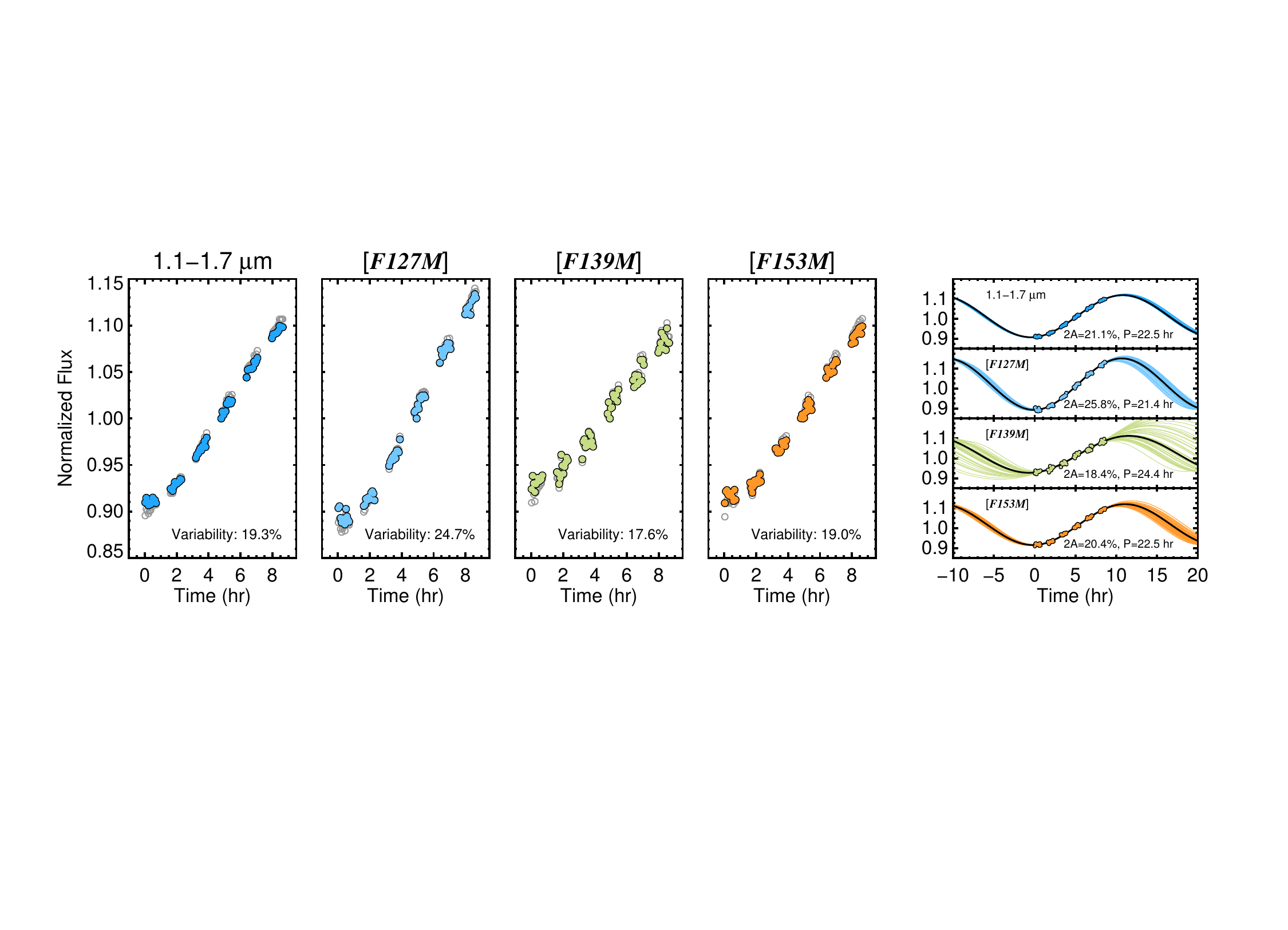}}
  \vskip -2 in
  \caption{\emph{HST} near-infrared spectroscopic light curves of VHS J1256--1257 b.
  Time-series photometry represent the full 1.1--1.68 $\mu$m region 
  as well as synthesized $F127M$, $F139M$, and $F153M$ bandpasses from left to right.
  Variability amplitudes
  are strongly wavelength dependent.  The highest amplitude (24.7\%) is seen in $F127M$ band with $F139M$
  and $F153M$ bands being lower (17.6\% and 19.0\%, respectively).  The spectroscopic light curve continues 
  to rise at the end of the observations so these values are likely lower limits for this epoch.  Sinusoidal fits to each 
  light curve are shown in the right-most panel.  50 random draws from the parameter posteriors are displayed.   \label{fig:lightcurves} } 
\end{figure*}

For the VHS~J1256--1257 b sequence, three spectra of VHS~J1256--1257 b, 
the host VHS~J1256--1257~AB (which is unresolved here), 
and the background star 2MASS~J12560179--1257390 
are extracted from each grism image (Figure~\ref{fig:grism1}). 
This results three spectroscopic light curves in units of both count rate (e$^{-}$ s$^{-1}$) and flux density (erg s$^{-1}$ cm$^{-2}$ $\mu$m$^{-1}$), as well as their corresponding measurement uncertainties.  
We then remove the ``ramp-effect'' systematics in the light curves (e.g., \citealt{Berta:2012ff}).
Here we use the Ramp Effect Charge Trapping Eliminator (RECTE; \citealt{Zhou:2017ir}) tool to model and correct these systematics. RECTE estimates the ``ramp-effect'' profile by assuming the systematics are introduced by charge trapping and delayed release due to detector defects. We adopt a set of pre-determined parameters that describe the charge trap numbers, efficiency, and trapping timescales. The remaining parameter that primarily determines the systematic profile is the pixel illumination level. We input the median-combined images to the RECTE model and calculate the ramp profile for each pixel, then sum the profiles of the same image columns to identify the correction term for each wavelength. Our final  spectroscopic light curves are obtained by dividing the correction term from the raw extracted spectra.

Light curves of VHS~J1256--1257 b, the host binary, and the background star are measured by integrating the spectra in the wavelength dimension. Four light curves are produced for each object: a broadband curve spanning 
1.10--1.68 $\mu$m where the $G141$ grism throughput is above 30\%, as well as 
time series photometry synthesized in the $F127M$ bandpass, $F139M$ bandpass, and $F153M$ bandpass (Figure~\ref{fig:lightcurves}).
The synthesized light curves are measured by convolving each spectrum with $HST$ transmission curves and then integrating the result.  Note that the red cutoff of the $G141$ grism truncates the traditional MKO $H$ bandpass by about 0.1~$\mu$m, so we chose to use medium-band \emph{HST} filters instead of standard near-infrared filters.

\subsubsection{GSC 6214--210~B and ROXs 42 B b}

\begin{figure*}
  \vskip 0.1 in
  \hskip 0 in
  \resizebox{7in}{!}{\includegraphics{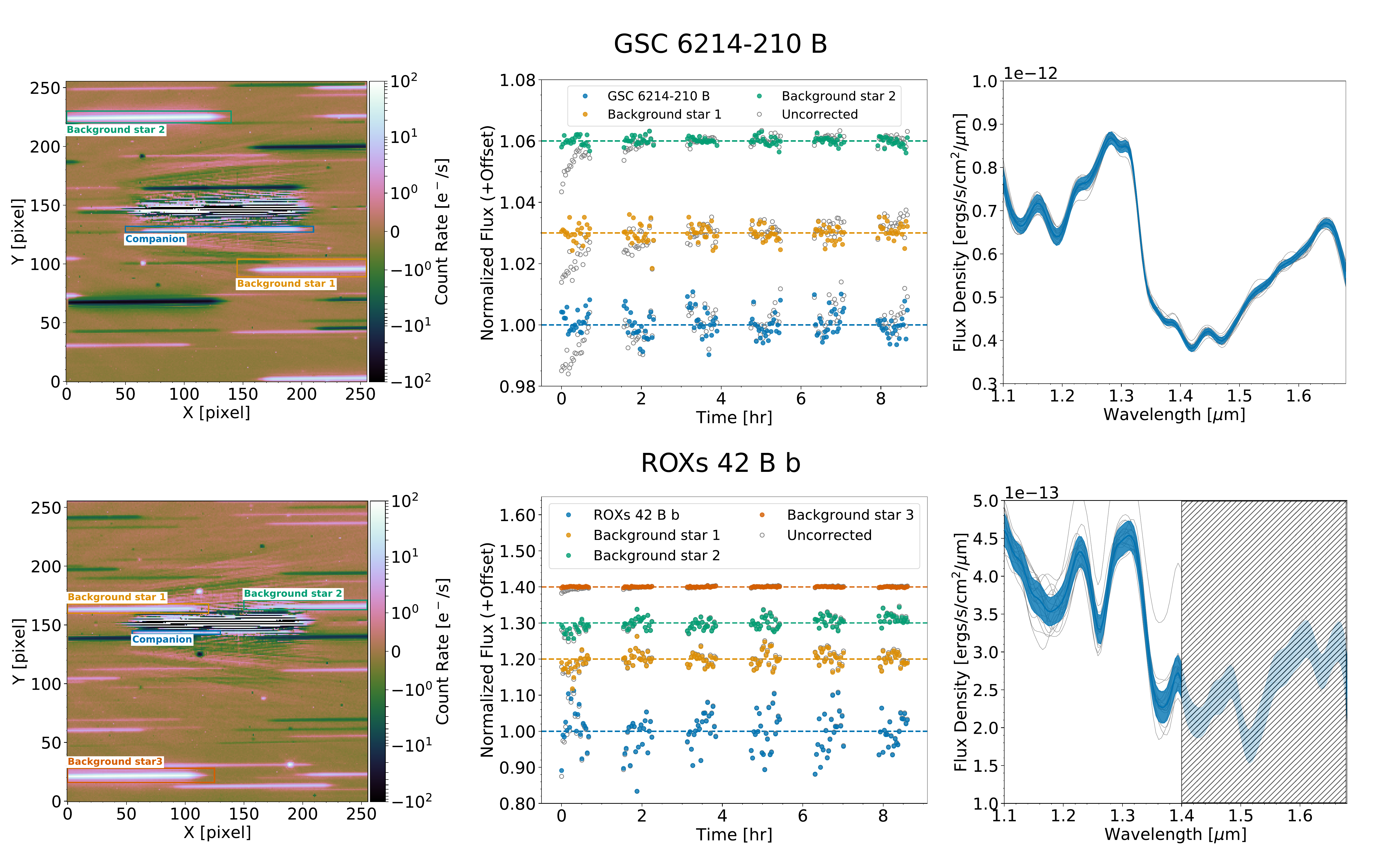}}
  \vskip -.2 in
  \caption{Results from $HST$ time-series grism observations of GSC~6214--210~B (top) 
  and ROXs~42~B~b (bottom).
  \emph{Left panels}: Grism images of each system.  PSF subtraction of the host star is carried out using the
  mirrored image about the central axis, which results in ``dipole spectra'' of the companion and background
  stars.  The dispersion direction is nearly aligned with
  the $x$-axis, with wavelength increasing to the right.
  \emph{Middle panels}: Extracted light curves of the companions and nearby background stars.
  Gray open circles represent the uncorrected photometry;
  colored symbols show the normalized flux (plus arbitrary offsets) 
  after correcting for detector ramp effects.
  No modulations are evident for GSC~6214-210~B ($<$1.2\%; 3$\sigma$ upper limit) nor ROXs 42 B b 
  ($<$15.6\%; 3$\sigma$ upper limit).
  \emph{Right panels}: Extracted spectra of each system.  For ROXs 42 B b, wavelengths beyond about
  1.4 $\mu$m are heavily contaminated by the host star and are not reliable.  
    \label{fig:grism2} }  
\end{figure*}

GSC 6214--210~B and ROXs 42 B b are both embedded in the PSF wings of their host stars in the grism images.  
For these observations,
our strategy is to subtract the spectral PSF of the primary star in the dispersion direction before spectral extraction.  
By assuming 
the spectral trace is symmetric with respect to the central axis, we mirror the image about this axis and subtract it individually from each science frame (Figure~\ref{fig:grism2}).
 Because the spectral trace is tilted by $\sim0.6\degr$ with respect to the image $x$-axis, 
 the mirrored image is first rotated  
 to match the original primary spectral PSF.  To optimize the subtraction, we set the template scaling parameter and the tilt angle as free parameters and fit them to minimize the rms subtraction residuals surrounding the companion spectral traces.  Following PSF subtraction, the steps to extract the spectra and correct the light curves for 
 these companions and several nearby background stars in the 
 images are identical to the procedure for VHS~J1256--1257 b.


\begin{figure}
  \vskip -0. in
  \hskip -3 in
  \resizebox{9.6in}{!}{\includegraphics{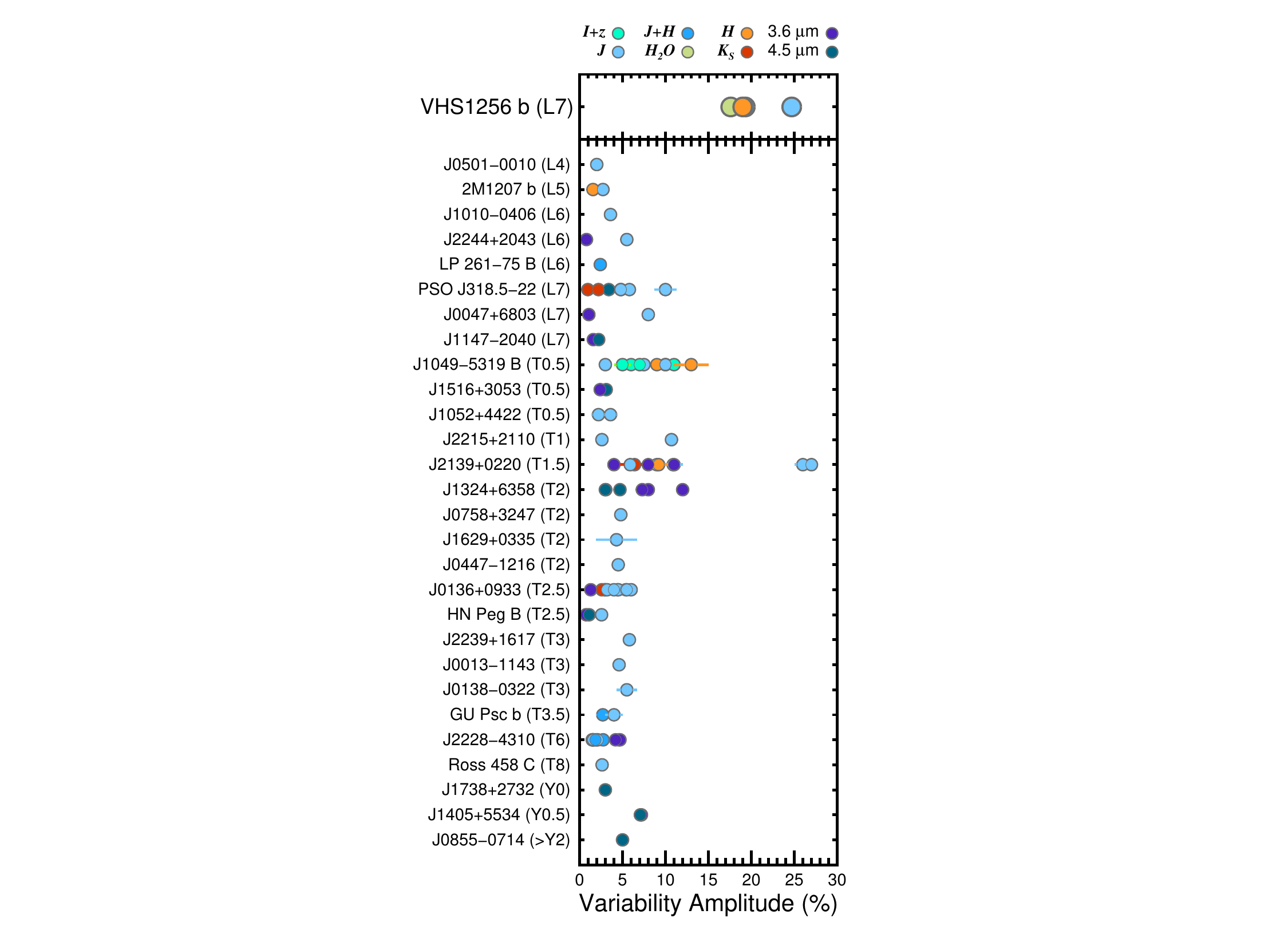}}
  \vskip -.1 in
  \caption{Amplitudes of the 28 strongest variable brown dwarfs and planetary-mass objects with 
  peak-to-peak brightness changes $>$2\% compared to our observations of VHS~J1256--1257 b (top panel).  
  Objects are sorted by spectral type and variability amplitudes are 
  color coded by filter.  VHS~J1256--1257 b has the second largest amplitude after 
  the T1.5 brown dwarf 2MASS~J21392676+0220226 (\citealt{Radigan:2012ki}; \citealt{Apai:2013fn}).
  Data are from \citet{Eriksson:2019aa},
   \citet[for LP-261-75~B]{Manjavacas:2017ie}, \citet[for GU~Psc~b]{Naud:2017do}, 
   and \citet[for GU~Psc~b]{Lew:2019aa}.  \label{fig:var} }
\end{figure}


\begin{figure}
  \vskip -0. in
  \hskip -.2 in
  \resizebox{3.7in}{!}{\includegraphics{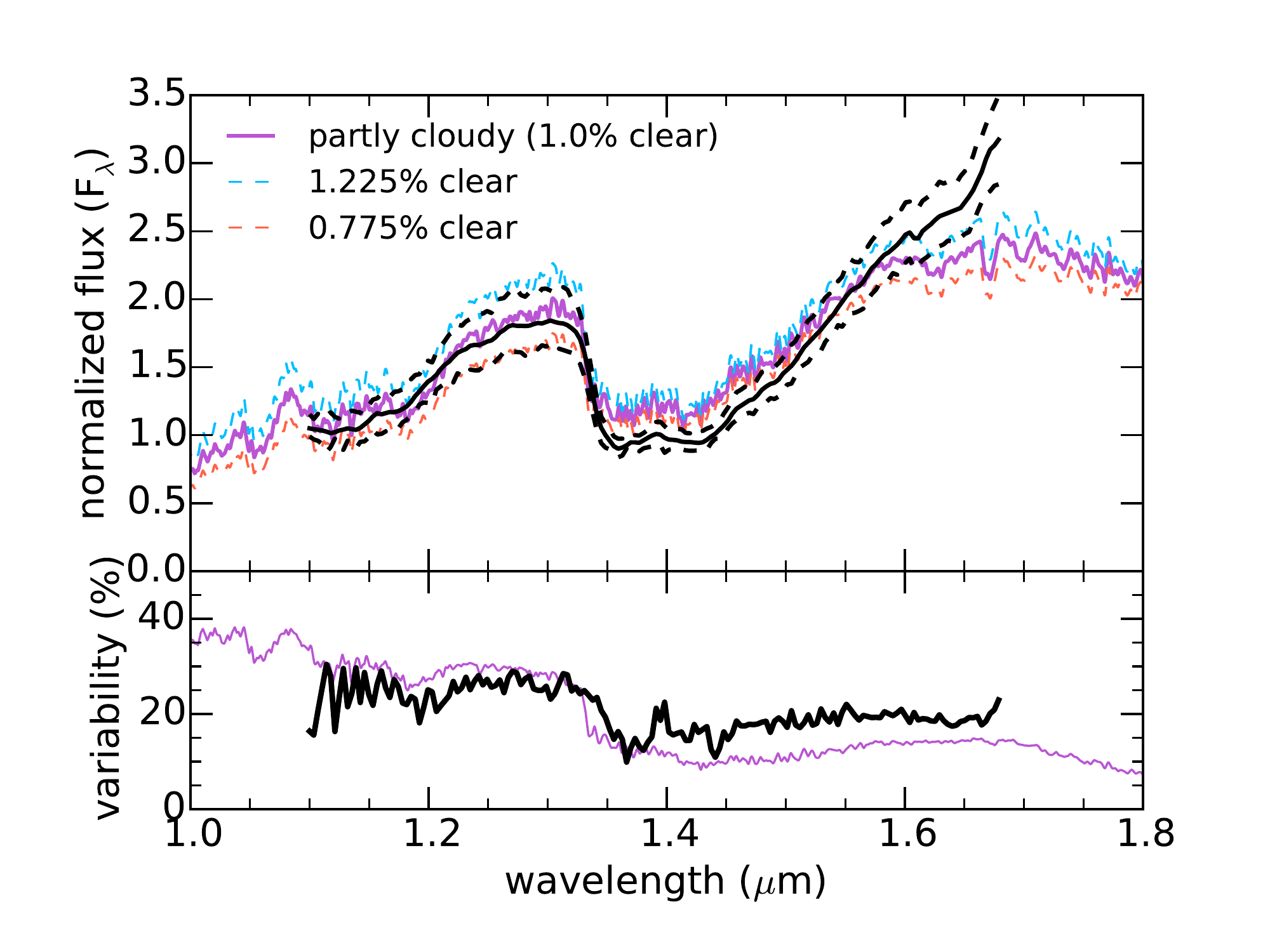}}
  \vskip -.1 in
  \caption{Simulated spectra used to represent atmospheric variability.
  By assuming the planet has partial cloud coverage,
  we are able to reproduce the emergent spectrum 
  (with model parameters 
  \Teff=900~K, log~$g$=4.0, and $f_\mathrm{sed}$=0.3; top panel), 
  the variability amplitude, and 
  the wavelength-dependent variability behavior (bottom panel) 
  using a 1\% clear surface with 0.225\% changes to the cloud-covering fraction.
  Dashed lines in the top panel show the minimum and maximum variability of the data (black) and
  model (orange and light blue).  \label{fig:model} }
\end{figure}

\section{Results and Discussion}{\label{sec:results}}

No variability is observed in the observations of 
GSC~6214-210~B and ROXs~42~B~b, but significant evolution is readily apparent for VHS~J1256--1257 b.
The brightness difference for VHS~J1256--1257 b is 
unusually large compared to the 28 strong variables with 
measured amplitudes greater than 2\% (Figure~\ref{fig:var}).
\citet{Metchev:2015dr} found that most brown dwarfs exhibit low-amplitude variability ($\approx$0.2--2\%),
but strong variability at the 2--10\% level is rarer, especially outside the L/T transition 
(\citealt{Radigan:2014fg}; \citealt{Eriksson:2019aa}).
Only five brown dwarfs are known to exhibit exceptionally high-amplitude variability above the 10\% level
(see \citealt{Eriksson:2019aa} for a recent compilation):
PSO~J318.5--22 (L7; \citealt{Biller:2018kd}; \citealt{Vos:2018aa}), 
WISE~J104915.57--531906.1 B (T0.5; \citealt{Gillon:2013um}; \citealt{Biller:2013jm}; \citealt{Buenzli:2015je})
2MASS~J22153705+2110554 (T1; \citealt{Eriksson:2019aa}), 
2MASS~J13243553+6358281 (T2; \citealt{Yang:2016dga}), and
2MASS~J21392676+0220226 (T1.5; e.g., \citealt{Radigan:2012ki}).
The brightness difference of 24.7\% we measure in the synthesized \emph{F127M} filter
for VHS~J1256--1257 b is the second-highest after 2MASS~J21392676+0220226,
which exhibits exceptionally strong variability up to 
27\% in $J$ band (\citealt{Apai:2013fn}).
Among these six strong variables (amplitudes $>$10\%), 
all have spectral types between L7 and T2 and 
three show spectroscopic and/or kinematic evidence of youth
(VHS~J1256--1257 b, PSO~J318.5--22, and 2MASS~J13243553+6358281).
These same three objects have masses near or within the planetary regime,
supporting trends of high amplitude modulation and low surface 
gravity found by \citet{Metchev:2015dr} and
\citet{Vos:2018aa}.

The large brightness changes from VHS~J1256--1257 b exhibit 
inflection points midway through the observations (Figure~\ref{fig:lightcurves})
which resemble sinusoidal modulations commonly observed 
among brown dwarfs and planetary-mass objects.
To estimate the rotation period and total variability amplitude, we fit the 
normalized broad-band light curve 
and synthesized light curves for individual bands assuming a 
sinusoidal model as follows: $f(t)$ = $a$ + $A \sin( (2\pi/P)t + \phi)$.\footnote{We tested 
both linear and sinusoidal relations to model the curves and calculated Bayesian Information Criterion 
(BIC) values, where BIC = $\chi^2$ + $k$ln$N$.  Here $k$ is the number of 
free parameters in the fit and $N$ is the number of data points (66).
The $\Delta$BIC between the two models is 1133 for the broadband light curve, 
indicating a very strong preference for the sine model.}
Here $f$ is the normalized flux,
$a$ is a constant offset, $A$ is the semi-amplitude, $P$ is the period, and $\phi$ is a phase offset.
Parameter posterior distributions are sampled 
using Markov Chain Monte Carlo with a Metropolis-Hastings algorithm.
Step sizes for trial values are chosen so that acceptance rates are between 20--30\%.
A uniform prior was adopted in $a$ (from 0--2), $A$ (from 0--2), $P$ (from 0--100 hours), and $\phi$ (from 0--2$\pi$).

Results of the sinusoidal fits are shown in Figure~\ref{fig:lightcurves}.
For the full 1.1--1.7~$\mu$m light curve, we infer a rotation period of 22.5$^{+0.3}_{-0.4}$~hr and
a peak-to-peak amplitude of 21.1$^{+0.2}_{-0.2}$\%.  The $\chi^2$ value of the best-fit curve is
126.1, the BIC value is 142.9, 
and the reduced $\chi^2$ value ($\chi^2_{\nu}$=$\chi^2$/($N$--$k$)) is 2.0.  Here $k$=4 and $N$=66. 
The synthetic photometry for individual bands implies similar periods 
but significant wavelength-dependent amplitude differences.  We find a period of
21.4$^{+0.8}_{-0.9}$~hr and a peak-to-peak amplitude of 25.8$^{+0.8}_{-0.7}$\% for $F127M$ 
($\chi^2$=65.1, BIC=81.9, $\chi^2_{\nu}$=1.05),
24$^{+4}_{-5}$~hr and an amplitude of 18.4$^{+2.2}_{-2.8}$\% for $F139M$ 
($\chi^2$=67.2, BIC=84.0, $\chi^2_{\nu}$=1.08),
and 22.5$^{+1.0}_{-0.9}$~hr and an amplitude of 20.4$^{+0.6}_{-0.7}$\% for $F153M$
($\chi^2$=81.4, BIC=98.2, $\chi^2_{\nu}$=1.23).
The implied rotation period of $\approx$21--24 hours assumes
the long-term variability is smooth and periodic.  It also neglects any possible latitudinal dependence of
the rotation rate, for example if the signal originates from non-equatorial features where banding
or differential rotation could produce a longer variability period than at equatorial regions.
These results highlight how unusual the properties of VHS~J1256--1257 b are
in terms of its variability amplitude, its strong chromatic modulations,
and its long rotation period compared to typical values of $\approx$3--20 hr 
for brown dwarfs (e.g., \citealt{Biller:2017kn}).
We note that this long rotation period can only be reconciled with the projected rotational velocity
of 13.5$^{+3.6}_{-4.1}$ km s$^{-1}$ from \cite{Bryan:2018hd} with a large physical radius and 
(nearly) edge-on viewing geometry.
The inferred inclination distribution of VHS~J1256--1257 b is explored in more detail in a companion study (Zhou et al., submitted).

Spectroscopic observations also enable 
searches for wavelength-dependent phase lags, which provide information about the nature and
distribution of surface features 
causing the brightness modulations.
These light curve phase shifts have been observed in several objects, especially
when comparing simultaneous near-infrared and mid-infrared light curves
(e.g., \citealt{Buenzli:2012gd}; \citealt{Biller:2018kd}), but this phenomenon is not universal among 
variable brown dwarfs.
We find phases of 4.73$^{+0.04}_{-0.04}$ rad, 
4.89$^{+0.15}_{-0.16}$ rad, and 4.77$^{+0.05}_{-0.05}$ rad
from our sinusoidal fits to the $F127M$, $F139M$, and $F153M$ light curves of VHS~J1256--1257 b, respectively.
The phase for the full 1.1--1.7~$\mu$m region is 4.84$^{+0.02}_{-0.02}$ rad.
We conclude that there is no significant evidence for distinct phase lags from these 
observations.

The observed variability of brown dwarfs has largely been attributed to the presence of clouds, 
which can produce a variety of time-dependent observable signatures.  This is expected to be especially
pronounced near the L-T transition as brown dwarfs cool and cloud decks sink below the photosphere at lower effective temperatures.
At higher altitudes (low pressures), patchy clouds tend to
produce gray modulations with little change in variability amplitude as a function of wavelength.
When located at lower altitudes (higher pressures), clouds can introduce a 
wavelength-dependence to the variability 
amplitude (e.g., \citealt{Buenzli:2012gd}).  In this scenario, spectral regions that probe deeper into the atmosphere 
(such as $J$ band) will be more susceptible 
to variability compared to wavelengths sensitive to higher altitudes (such as the 1.4~$\mu$m water band).
These wavelength-dependent amplitude differences are therefore generally interpreted as signatures of
mid-altitude clouds and have predominantly been observed in early T dwarfs, whereas 
gray (wavelength-independent) variability likely arises from high-altitude clouds and appears to be more common 
among mid-L dwarfs (e.g., \citealt{Yang:2015ji}; \citealt{Artigau:2018kg}).
VHS~J1256--1257 b has a spectral type of L7 but exhibits unusually strong \emph{relative} spectral variability
of 5.7\% between $F127M$ and $F153M$, and
even larger differences of 7.1\% between $F127M$ and $F139M$. 
Aside from a small increase at $\approx$1.39~$\mu$m in the middle of the water band,
these changes are relatively constant within each 
bandpass (Figure~\ref{fig:grism1}).  
In this respect VHS~J1256--1257 b more closely resembles the spectral variability behavior of early T dwarfs
than mid-L dwarfs.

VHS~J1256--1257 b also shares similarities 
with the two other low-mass variable objects WISE~J0047+6803 and PSO~J318.5--22: 
all have comparable spectral types (L6--L7),
extremely red near-infrared 
slopes ($J$--$K_S$=2.5--2.8~mag), and similar wavelength-dependent variability 
patterns from 1.1--1.7~$\mu$m with the strongest variability at $J$ band and the weakest
variability in the 1.4~$\mu$m water band (\citealt{BenWPLew:2016kg}; \citealt{Biller:2018kd}).  
However, these traits are all significantly enhanced in
VHS~J1256--1257 b, which may be a result of thinner high-altitude clouds or distinct dust properties  
compared to the other two objects.

To provide physical insight into the observed variability, 
partly cloudy model spectra were generated following the methodology presented 
in \citet{Marley:2010kx} and \citet{Morley:2014dn}. 
All models use \Teff=900 K and log g=4.0 and assume that the 
atmosphere is in chemical and radiative-convective equilibrium. A small grid of sedimentation 
efficiency (\fsed) and cloud-free surface fraction (\fhole) was considered (\fsed=0.3--2; \fhole=1--10\%); 
very lofted clouds (\fsed=0.3) with a small cloud-free surface fraction (1\%) provided the best match, 
as expected for this red, dusty object. After converging this partly-cloudy model, moderate resolution spectra through the cloudy and clear columns were generated. The trend and amplitude in variability can be matched by varying the cloud-covering fraction by 0.225\% (from 0.775\% cloud-free to 1.225\% cloud-free). 

The resulting model is meant to demonstrate that the overall trend---higher amplitude variability in $J$ band 
relative to the water band and $H$ band---and large amplitude can be matched by 
considering non-uniform clouds. Future work should investigate disequilibrium chemistry 
(a likely cause of the model--data discrepancy in $H$ band); 
provide detailed fits to larger grids of model spectra; 
and investigate different cloud treatments (e.g., thicker and thinner clouds and latitudinal banding
rather than thick clouds and cloud-free regions)---particularly using two- or three-dimensional models. 

\subsection{Upper limits on variability for GSC~6214-210~B and ROXs~42~B~b}

In contrast, our light curves of GSC~6214-210~B and ROXs 42 B b appear to be flat with
no evidence of significant variability.  The photometric precision for these companions is limited by the PSF subtraction
in the grism images.  This procedure was substantially more challenging for ROXs 42 B b owing to the close
separation of its host star (1$\farcs$2).  We measure rms levels of 0.4\% for GSC~6214-210~B, or a 3$\sigma$ upper limit 
of 1.2\%.  For ROXs 42 B b we find an rms of 5.2\%, or a 3$\sigma$ upper limit 
of 15.6\%.  

\section{Summary}{\label{sec:summary}}

We have obtained \emph{HST}/WFC3 near-infrared 
time-series spectroscopy of the three young companions
VHS~J1256--1257 b, GSC~6214--210~B, and ROXs~42~B~b
all of which have masses at or near the deuterium-burning boundary.
Results from this program are summarized below:

\begin{enumerate}
\item Our 8.5~hr observations show strong brightness changes in VHS~J1256--1257 b
at the level of 19.3\% across the full 1.1--1.68~$\mu$m region. For the synthesized
bandpasses we find 24.7\% variability in $F127M$, 17.6\% variability in
$F139M$, and 19.0\% variability in $F153M$.
These are the highest variability measurements of any substellar object
after the T1.5 brown dwarf 2MASS~J21392676+0220226.  
The evolving light curve of VHS~J1256--1257 b was still rising at the end of the
observations, implying the full variability amplitude is likely to be even larger at this epoch.
\item We interpret these substantial wavelength-dependent variability amplitudes as evidence
for mid-altitude cloud decks, similar to (but much stronger than) 
the other young red L dwarfs WISE~J0047+6803 and PSO~J318.5--22.
The spectral shape, large-amplitude brightness modulations, and wavelength-dependent variability
can be reproduced with atmospheric models comprising a 1\% clear surface coupled 
with 0.225\% changes in cloud-covering fraction.
\item Sinusoidal fits to the light curves of VHS~J1256--1257 b imply rotation periods 
between 21--24~hr 
for the full 1.1--1.68~$\mu$m spectral region
and individual synthesized bands.
We find no evidence of phase shifts from these observations, although 
only about 30--40\% of a single rotation period appears to have been sampled.
\item No variability is observed from GSC~6214--210~B or ROXs~42~B~b
at the $<$1.2\% and $<$15.6\% levels (3$\sigma$ upper limits), respectively.  These precisions are limited
by the close separations of these companions to their host stars (1$\farcs$2--2$\farcs$2).

\end{enumerate}

The large separation of VHS~J1256--1257 b from its host star (8$\farcs$1) makes it an excellent
target for follow-up observations spanning longer time baselines, broader wavelength coverage,
and higher spectral resolution.  The high near-infrared amplitude for this object also means that 
it especially amenable to ground-based photometric monitoring.
Altogether these results for VHS~J1256--1257 b 
suggest that other young planetary-mass companions at the L/T transition 
like HR~8799~bcde, HD~95086~b, HIP~65426~b, and 2MASS~J22362452+4751425~b
are promising companions to search for variability.

\acknowledgements

The authors thank Heather Knutson and Trent Dupuy for helpful discussions.
This research is based on observations made with the NASA/ESA \emph{Hubble Space Telescope} obtained from the Space Telescope Science Institute, which is operated by the Association of Universities for Research in Astronomy, Inc., under NASA contract NAS 5-26555. These observations are associated with program 15197.
B.P.B. acknowledges support from the National Science Foundation grant AST-1909209.

\facility{HST (WFC3)}


\end{document}